\documentclass[12pt]{article}
\usepackage{euscript,amssymb}
\def\bc{\begin{center}}
\def\ec{\end{center}}
\def\bu{\begin{enumerate}}
\def\eu{\end{enumerate}}
\def\p1leiste #1 #2 #3 {{\noindent\Large Volume #1 \hfill {#2}
\hfill {Number #3} \vspace{2mm}\par\hrule\par\vspace{8mm}}}
\def\an #1 #2 #3 #4 #5 {{\noindent\small Astron. Nachr. #1 (#2) #3, #4-#5
\bigskip\par}}
\def\address #1END {{\vspace{9mm}\noindent\small Address of the author:
 \medskip \\ #1}}
\def\min #1 #2{\mbox{$#1_{\cdot}^{'}#2$}}
\def\mg #1 #2{\mbox{$#1_{\:\cdot}^{m}#2$}}

\title{Generalized de Sitter solution in multidimensional cosmology
with static internal spaces}
\author{ A.Zhuk \\ Fachbereich Physik, Freie Universit\"at Berlin}

\date{
Astron. Nachr. 316 (1995) 269 - 274}
\begin{document}
\maketitle
\abstract{A multidimensional cosmological model with space-time
consisting of $n (n\geq 2)$ Einstein spaces $M_i$ is investigated
in the presence of a cosmological constant $\Lambda$ and a
homogeneous minimally coupled free scalar field. Generalized de
Sitter solution was found for $\Lambda >0$ and Ricci-flat external
space for the case of static internal spaces with fine tuning of
parameters. }

PACS numbers: 04.50.+h, 98.80.Hw

\vspace{2cm}

{\bf{1 INTRODUCTION}}

\vspace{0.5cm}

 The de Sitter solution (De Sitter (1917a, 1917b)) in
4-dimensional space-time is remarkable from different points of
view. Firstly, it describes the unique curved space-time with
maximum of symmetry (Weinberg (1972), Misner, Thorne, and Wheeler
(1974), Hawking, Ellis (1973), Birrell, and Davies (1982)) .
Secondly, this solution simulates the inflationary stages of the
evolution of universe (Guth (1981), Linde (1984)). From the other
hand it is well known that modern theories of unified physical
interactions take place in multidimensional spaces only. Thus, it
will be important to generalize the de Sitter solution for
multidimensional cosmological models (MCM). In all MCM a mechanism
of dimensional reduction (compactification) of extra dimensions
should be present. One of the possibilities to solve this problem
consists in the proposal that all extra dimensions are static and
small (a few orders of Planck's length) from the very beginning.
In spite of smallness of extra dimensions the iternal spaces have
strong influence on the parameters of external space, for example,
on the rate of the evolution of external space.

In this paper we shall find the generalization of the de Sitter solution in
the case of MCM with static internal spaces where
internal spaces are Einstein ones.

\vspace{1cm}

{\bf 2 THE MODEL}

\vspace{0.5cm}

We consider a cosmological model with $n(n>1)$ Einstein spaces containing a
massless minimally coupled free scalar field and a positive
cosmological constant $\Lambda$. The metric of the model
\begin{equation}
g = - \exp{\left[2\gamma (\tau)\right]}\, d\tau \otimes d\tau +
\sum_{i=1}^{n}\exp{\left[2\beta^i(\tau)\right]}g_{(i)}
\end{equation}                     
is defined on the manifold
\begin{equation}
M=R\times M_1\times\cdots\times M_n ,
\end{equation}                     
where the manifold $M_i$ with the metric $g_{(i)}$ is an Einstein
space of dimension $d_i$, i.e.
\begin{equation}
R_{m_i n_i}\left[g_{(i)}\right]=\lambda_i g_{(i)m_i n_i}
\end{equation}                     
$i=1,\ldots,n; n\geq 2.$ In the particular case of the spaces of
constant curvature $\lambda_i=k_i(d_i-1)$, where $k_i=\pm 1,0$. The
total dimension of the space-time $M$ is $D=1+\sum_{i=1}^n d_i$.

The action of the model is
\begin{equation}
S=\frac{1}{2}\int d^D x \sqrt{|g|}\left\{R[g]-
{\left(\nabla\varphi\right)}^2 - 2\Lambda\right\} + S_{GH},
\end{equation}
where $R[g]$ is the scalar curvature of the metric $g$ and $S_{GH}$
is the standard Gibbons-Hawking boundary term
(Gibbons, and Hawking (1977))
. This type of
MCM (without scalar field and cosmological constant) was considered
first in
(Ivashchuk, Melnikov, and Zhuk (1989))
. Some integrable cases were investigated in
(Zhuk (1992a),  Zhuk (1992b),
Bleyer, Ivashchuk, Melnikov, and Zhuk (1994))
. Following paper
(Ivashchuk, Melnikov, and Zhuk (1989))
we get that the field
equations, corresponding to the action (4), for the metric (1) in the
harmonic time gauge $\gamma=\sum_{i=1}^n d_i \beta^i$ are equivalent
to the Lagrange equations, corresponding to the Lagrangian
\begin{equation}
L=\frac{1}{2}\left(\sum_{i,j=1}^n G_{ij}{\dot \beta}^i
{\dot \beta}^j + {\dot \varphi}^2 \right) - V
\end{equation}                                 
with the energy constraint imposed
\begin{equation}
E=\frac{1}{2}\left(\sum_{i,j=1}^n G_{ij}{\dot \beta}^i
{\dot \beta}^j + {\dot \varphi}^2 \right) + V = 0
\end{equation}                            

Here, the overdot denotes differentiation with respect to the
harmonic time $\tau$. The components of minisuperspace  metric read
$G_{ij}=d_i\delta _{ij} - d_i d_j$ and the potential is given by
\begin{equation}
V=\exp {\left(2\sum_{i=1}^n d_i \beta ^i\right)}\left[-\frac{1}{2}
\sum_{i=1}^n d_i \lambda _i e^{-2\beta ^i} + \Lambda \right].
\end{equation}                            

In our paper we consider MCM with static internal spaces.
Let the factor space $M_1$ be dynamical external
space.  All the other factor spaces $M_i\, (i=2,\ldots ,n)$ are
considered as internal and static. They should be compact and the
internal dimensions have the size of order of Planck's length
$L_{PL}\sim {10}^{-33}cm$. The scale factors of the internal factor
spaces should be constant: $a_i=e^{\beta ^i}\equiv a_{(0)i}\quad
(i=2,\ldots ,n)$. The problem of compactification with static internal
spaces for the
models with the topology (2) was considered in
(Bleyer and Zhuk (1994))
. According
to the theorems proved there the static compactification for our
model takes place only in the case of fine tuning of the parameters:
\begin{equation}
\frac{\lambda _i}{a_{(0)i}^2} = \frac{2\Lambda}{D-2}\quad,\qquad
i=2,\ldots ,n\,.
\end{equation}                            
It follows from (8) that for $\Lambda >0$ parameters $\lambda _i$
should be positive also. If $M_i\, (i=2,\ldots ,n)$ are spaces of
constant curvature, they should have positive constant curvature.

Let us investigate the model where our external space $M_1$ is
Ricci-flat, i.e. $\lambda _1=0$. Then the constraint  (6) has form
\begin{equation}
d_1(d_1-1){\left({\dot \beta}^1\right)}^2 = \nu ^2 + e^{2d_1\beta ^1}
\frac{2(d_1-1)\Lambda}{D-2}\prod_{k=2}^n a_{(0)k}^{2d_k},
\end{equation}                            
where $\dot \varphi = \nu = const$ is the first integral of the
equation $\ddot \varphi = 0$. Parameter $\nu ^2$ plays the role of
energy (Zhuk (1992b)).\\

\vspace{1cm}
{\bf 3  GENERALIZED DE SITTER SOLUTION}

\vspace{0.5cm}

We can rewrite the equation (9)
as follows
\begin{equation}
{\left({\dot \beta}^1\right)}^2 = {\tilde \nu}^2 + {\tilde \Lambda}
e^{2d_1\beta ^1},
\end{equation}                            
where the constants are
\begin{equation}
{\tilde \nu}^2 = \frac{\nu ^2}{d_1(d_1-1)}
\end{equation}                            
and
\begin{equation}
\tilde \Lambda = \frac{2\Lambda}{d_1(D-2)}\prod_{k=2}^n
a_{(0)k}^{2d_k}.
\end{equation}                            
It is clear from (10) that the dynamical behaviour of the scale
factor $a_1=\exp {\beta ^1}$ depends on the sign of $\nu ^2$.\\

a) {\bf $\nu ^2>0$: real scalar field.}\\
The solution of equation (10) has the form
\begin{equation}
a_1(\tau) = {\left[\left.{\tilde \nu}^2\right/\tilde \Lambda
\right]}^{1/2d_1}{\left|\sinh {\tilde \nu d_1(\tau - \tau _0)}\right|}
^{-1/d_1}\,,\quad -\infty<\tau <+\infty
\end{equation}                            
where $\tau_0$ is the constant of integration.\\

The synchronous time $t$ and the harmonic time $\tau$ are connected
by the differential equation
(Ivashchuk, Melnikov, and Zhuk (1989))
\begin{equation}
e^{\gamma (\tau)}d\tau = dt
\end{equation}                            
where $\gamma (\tau) = \sum_{i=1}^n d_i \beta ^i$. It is not difficult
to get the connection
$$
{\left[\sinh {\tilde \nu d_1(\tau - \tau _0)}\right]}^{-1} =
\sinh {\sqrt {\left.2\Lambda d_1\right/\left(\sum_{k=1}^n
d_k-1\right)}(t - t_0)}
$$
With the help of this connection we obtain the expression for the
scale factor $a_1$ with respect to the synchronous time
\begin{equation}
a_1(t) = {\left[\left.{\tilde \nu}^2\right/\tilde \Lambda\right]}^{1/2d_1}
{\left|\sinh {\sqrt {\left.2\Lambda d_1\right/\left(\sum_{k=1}^n
d_k-1\right)}(t - t_0)}\right|}^{1/d_1}\,,\quad -\infty< t <+\infty
\end{equation}                            

b) {\bf $\nu ^2<0$: imaginary scalar field.}\\
The solution of equation (10) in this case is
\begin{equation}
a_1(\tau) = {\left[\left.\left|{\tilde
\nu}^2\right|\right/\tilde \Lambda\right]}^{1/d_1}\left\{
\cos {{\left|{\tilde \nu}^2\right|}^{1/2}d_1(\tau - \tau
_0)}\right\}^{-1/d_1}\,,\quad |\tau - \tau _0|\leq \left.\pi \right/
\left(2d_1{\left|{\tilde \nu}^2\right|}^{1/2}\right)
\end{equation}                            
The connection between harmonic and synchronous times is
\begin{equation}
\cos {{\left|{\tilde \nu}^2\right|}^{1/2}d_1(\tau - \tau _0)} =
{\left[\cosh {\sqrt {\left.2\Lambda d_1\right/\left(\sum_{k=1}^n
d_k - 1\right)}(t-t_0)}\right]}^{-1}
\end{equation}                            
In the synchronous time $t$ we have the solution
\begin{equation}
a_1(t) = {\left[\left.\left|{\tilde
\nu}^2\right|\right/\tilde \Lambda\right]}^{1/2d_1}
{\left\{\cosh {\sqrt {\left.2\Lambda d_1\right/\left(\sum_{k=1}^n
d_k - 1\right)}(t-t_0)}\right\}}^{1/d_1},
\end{equation}                            
$$-\infty< t <+\infty$$
It is clear that for $\nu ^2<0$ the solution of the Euclidean analog
of the equation (10) exists also. It means on quantum level that the
transitions with changing of the topology take place in this case,
for example, birth from nothing (Vilenkin (1983)).\\

c) {\bf $\nu ^2=0$: absence of scalar field.}\\
In this case the solution of the equation (10) in the harmonic time
$\tau$ reads
\begin{equation}
a_1(\tau) = {\left[d_1^2{\tilde \Lambda}\right]}^{-1/2d_1}
{\left|\tau - \tau _0\right|}^{-1/d_1}\,,\quad -\infty< \tau <+\infty
\end{equation}                            
The harmonic and synchronous times are connected by formula
\begin{equation}
\left|\tau - \tau _0\right| = \exp {\left[\pm \sqrt {\left.2\Lambda
d_1\right/\left(\sum_{k=1}^n
d_k - 1\right)}(t-t_0)\right]}
\end{equation}                            
The solution (19) with the help of the expression (20) may be
rewritten in the form
\begin{eqnarray}
a_1(t) & = & {\left[d_1^2{\tilde \Lambda}\right]}^{-1/2d_1}
 \exp {\left[\pm \sqrt {\left.2\Lambda
\right/d_1\left(\sum_{k=1}^n
d_k - 1\right)}(t-t_0)\right]} \equiv  \nonumber \\
& \equiv &  a_{(0)1} \exp {\left[\pm \sqrt {\left.2\Lambda
\right/d_1\left(\sum_{k=1}^n
d_k - 1\right)}t\right]}
\end{eqnarray}                            
where $t_0$ and $a_{(0)1}$ are constants of integration. It is easy
to see that the solutions (15) and (18) tend asymptotically to (21)
when $|t|\rightarrow \infty$.

The formula (21) (with the sign $"+"$) represents a multidemensional
generalization of the de Sitter solution. If we put formally $n=1,
d_1=3$ in (21) we get the usual form of the scale factor in
4-dimensional de Sitter space-time: $a\sim \exp {\left(\sqrt {\Lambda
/3}t\right)}$. In the case $d_1=3, n>1$ the higher dimensions have an
imprint in the external space $M_1$ through the exponent: $\Lambda /3
\rightarrow 2\Lambda /\left[3\left(\sum_{k=2}^n d_k+2\right)\right]$.
We would like to stress once more, that $\Lambda$ and the parameters
of the internal spaces are fine tuned according to equation (8).

To show more explicitly that (21) represent a multidimensional
generalization of the de Sitter solution, let us consider $M_1$ as a
$d_1$-dimensional flat space with the metric $g_{(1)}=\sum_{i=1}^{d_1}
dx^i\otimes dx^i$. Then, the multidimensional metric (1) in
synchronous system reads
\begin{equation}
g=-dt\otimes dt + a_1^2(t)\sum_{i=1}^{d_1}dx^i\otimes dx^i +
\sum_{i=2}^n a_{(0)i}^2g_{(i)}
\end{equation}                            
where
\begin{equation}
a_1(t)=\exp {Ht}
\end{equation}                            
We introduced the Hubble constant $H=\sqrt {\left.2\Lambda \right/d_1
\left(\sum_{k=1}^n d_k-1\right)}$ and without loss of generality we
have put $a_{(0)1}=1$.

The coordinates transformation
\begin{eqnarray}
y^0 & = & \frac{1}{H}\sinh {Ht} + \frac{H}{2}e^{Ht}{|\vec x|}^2
\nonumber \\
y^{d_1+1} & = &\frac{1}{H}\cosh {Ht} -\frac{H}{2}e^{Ht}{|\vec x|}^2  \\
y^i & = & e^{Ht}x^i\,,\quad i=1,\ldots ,d_1   \nonumber
\end{eqnarray}
gives $(D+1)$-dimensional metric
\begin{equation}
g=-dy^0\otimes dy^0 + \sum_{i=1}^{d_1+1}dy^i\otimes dy^i +
\sum_{i=2}^n a_{(0)i}^2g_{(i)}
\end{equation}                            
where the coordinates $y$ satisfy the equation of the
$(d_1+1)$-dimensional hyperboloid
\begin{equation}
-{(y^0)}^2+\sum_{i=1}^{d_1+1}{(y^i)}^2=1/H^2
\end{equation}                            
Thus, the generaralized de Sitter solution is the hypersurface
${\cal H}^{d_1+1}\times M_2\times \cdots \times M_n$ in the
$(D+1)$-dimensional space with the topology $M^{d_1+2}\times
M_2\times \cdots \times M_n$ when ${\cal H}^{d_1+1}$ is
$(d_1+1)$-dimensional hyperboloid, $M^{d_1+2}$ is
$(d_1+2)$-dimensional Minkowski space and $M_i\, (i=2,\ldots ,n)$ are
freezed compact Einstein spaces.

The solution (22), (23) is the metric in representation of the
stationary universe
(Weinberg (1972), Misner, Thorne, and Wheeler (1974),
Hawking, Ellis (1973), Birrell, and Davies (1982))
. We can choose another section of the
hyperboloid if we take new coordinates $(\bar t, \theta ^1,\ldots ,
\theta ^{d_1})$:
\begin{eqnarray}
y^0 &=& \frac{1}{H}\sinh {H\bar t} \nonumber\\
y^1 &=& \frac{1}{H}\cosh {H\bar t}\cos {\theta ^1} \nonumber\\
y^2 &=& \frac{1}{H}\cosh {H\bar t}\sin {\theta ^1}\cos
{\theta ^2} \nonumber\\
y^3 &=& \frac{1}{H}\cosh {H\bar t}\sin {\theta ^1}\sin
{\theta ^2}\cos {\theta ^3}\\
\vdots \nonumber\\
y^{d_1} &=& \frac{1}{H}\cosh {H\bar t}\sin {\theta ^1}\sin
{\theta ^2}\ldots \cos {\theta ^{d_1}} \nonumber\\
y^{d_1+1} &=& \frac{1}{H}\cosh {H\bar t}\sin {\theta ^1}\sin
{\theta ^2}\ldots \sin {\theta ^{d_1}} \nonumber
\end{eqnarray}
In this coordinate system the metric (25) reads
\begin{equation}
g=-d\bar t\otimes d\bar t + \frac{1}{H^2}\cosh^2 {H\bar t}\ \bar g_{(1)}+
\sum_{i=2}^n a_{(0)i}^2g_{(i)}
\end{equation}                            
where $\bar g_{(1)}$ is the metric on $S^{d_1}$:
\begin{equation}
\bar g_1=d\theta ^1\otimes d\theta ^1+\sin ^2{\theta ^1}d\theta
^2\otimes d\theta ^2+\cdots +\sin ^2{\theta ^1}\ldots
\sin ^2{\theta ^{d_1-1}}d\theta ^{d_1}\otimes d\theta ^{d_1},
\end{equation}
$$0\leq \theta ^1,\ldots ,\theta ^{d_1-1}\leq \pi\,,\quad
0\leq \theta ^{d_1}\leq 2\pi.$$
Thus, for this section of the hyperboloid ${\cal H}^{d_1+1}$ the factor
space $M_1$ has positive constant curvature. It is well known
property of the de Sitter space (Hawking, Ellis (1973),
Birrell, and Davies (1982)).

\vspace{1cm}
{\bf 4 CONCLUSION}

\vspace{0.5cm}

In this paper we investigated multidimensional cosmological models
with $n(n>1)$ Einstein spaces in the presence of the cosmological
constant $\Lambda$ and a homogeneous minimally coupled free scalar
field as a matter source. Generalized de Sitter solution was found
for positive cosmological constant and Ricci-flat external space
for the case of static internal spaces with fine tuning of
parameters. All internal spaces have positive curvature.

The solutions obtained here give us an interesting example of the
development of Einstein's idea concerning $\Lambda$-term. Although
Einstein considered his original assumption the "bigest blunder" of
his life, in multidimensional theories this idea has new important
features. We saw here that the cosmological constant plays the role
of a "double agent". From one side, it serves to keep the internal
factor spaces $M_i\,(i=2,\ldots ,n)$ freezed through the fine tuning of
parameters (see equation (8)) in the spirit of Einstein idea. From
other side, the cosmological constant provides the de Sitter-like
behaviour for the external space.

We would like to note that the solutions with exponential behaviour
of the scale factors were found also in
(Bleyer, Ivashchuk, Melnikov, and Zhuk (1994))
for the model with
all Ricci-flat factor spaces and positive cosmological constant.
However, the isotropization condition for all dimensions takes place
in this model and there is no compactification of the internal
spaces. For $\Lambda <0$ classical as well as quantum wormhole
solutions were found there. Classical wormhole solutions were
obtained also in this paper if $\Lambda <0$ and only one of the $M_i$
being Ricci-flat for the case of freezed
internal dimensions with fine tuning parameters similar to the
equation (8). In this case all internal spaces should have negative
curvature.

{\bf Acknowledgement.} The work was sponsored by DFG grant
436-UKR-17-30-93. I wish to thank also Professor Dr. H.Kleinert and
the institute for Theoretical Physics at the Free University of
Berlin for their hospitality during the preparation of the paper.


\bigskip
\newpage
%
\begin{list}{}{\itemsep=0pt\parsep=0pt
\labelsep=10pt\leftmargin=10pt\labelwidth=0pt}
\small
\item
[Birrell, N.D., and Davies, P.C.W.:] 1982, Quantum Fields in Curved Space
(Cambridge University Press, Cambridge).
\item
[Bleyer, U., Ivashchuk, V.D., Melnikov V.N., and Zhuk, A.:] 1994,
Nucl. Phys., {\bf B429}, 177.
\item
[Bleyer, U., and Zhuk, A.:] 1994, Class. Quant. Grav., {\bf 11}, 1.
\item
[De Sitter, W.:] 1917a, Proc. Kon. Ned. Akad. Wet., {\bf 19}, 1217.
\item
[De Sitter, W.:] 1917b, Proc. Kon. Ned. Akad. Wet., {\bf 20}, 229.
\item
[Gibbons, G.W., and Hawking, S.W.:] 1977, Phys. Rev., {\bf D15}, 2752.
\item
[Guth, A.H.:] 1981, Phys. Rev., {\bf D23}, 347.
\item
[Hawking, S.W., Ellis, G.F.R.:] 1973, The Large Scale Structure of
Space-Time (Cambridge University Press, Cambridge).
\item
[Ivashchuk, V.D., Melnikov, V.N., and Zhuk, A.I.:] 1989, Nuovo
Cimento, {\bf B104}, 575.
\item
[Linde, A.D.:] 1984, Rep. Prog. Phys., {\bf 47}, 925.
\item
[Misner, C.W., Thorne, K.S., and Wheeler, J.A.:] 1974, Gravitation
(Freeman, San Francisco).
\item
[Vilenkin, A.:] 1983, Phys. Rev., {\bf D27}, 2848.
\item
[Weinberg, S.:] 1972, Gravitation and Cosmology: Principles and
Applications of the General Theory of Relativity (Wiley, New York).
\item
[Zhuk, A.:] 1992a, Phys. Rev., {\bf D45}, 1192.
\item
[Zhuk, A.:] 1992b, Class. Quant. Grav., {\bf 9}, 2029.
\end{list}
\address
A. Zhuk\\
Fachbereich Physik, Freie Universit\"at Berlin\\
Arnimallee 14\\
D-14195, Germany\\
e-mail: zhuk@einstein.physik.fu-berlin.de \\ \\
Permanent address:\\
Department of Physics,\\
University of Odessa, \\
2 Petra Velikogo,\\
Odessa 65100, Ukraine\\
e-mail: zhuk@paco.net END
\end{document}